\documentclass[12pt]{iopart}
\usepackage{epsfig}
\begin{document}

\title{Aging in Spin Glasses in three, four and infinite dimensions}

\author{
S. Jim\'enez\dag\ddag, V. Mart\'{\i}n-Mayor\P\ddag, G. Parisi\S\ and 
A. Taranc\'on\dag\ddag}

\address{\dag\ Departamento de F\'{\i}sica Te\'orica,
Facultad de Ciencias, 
Universidad de Zaragoza, 50009 Zaragoza, Spain.}
\address{\ddag\ Instituto de Biocomputaci\'on y  F\'{\i}sica de
Sistemas Complejos (BIFI),
Facultad de Ciencias, Universidad de Zaragoza, 50009 Zaragoza, Spain.}
\address{\P\  Departamento de F\'{\i}sica Te\'orica I, Facultad de C.C. 
F\'{\i}sicas, 
Universidad Complutense de Madrid, 28040 Madrid, Spain.}
\address{\S\  Dipartimento di Fisica, , Sezione INFN, SMC and UdRm1 of INFM, Universit\`a di Roma {\em La Sapienza},
P.le A. Moro 2, Roma I-00185, Italy.}

\eads{\mailto{sergio@rtn.unizar.es}, \mailto{victor@lattice.fis.ucm.es}, \mailto{Giorgio.Parisi@roma1.infn.it},\mailto{tarancon@sol.unizar.es}}

\date{\today}

\begin{abstract}
The SUE machine is used to extend by a factor of 1000 the time-scale
of previous studies of the aging, out-of-equilibrium dynamics of the
Edwards-Anderson model with binary couplings, on large lattices
($L=60$). The correlation function, $C(t+t_w,t_w)$, $t_w$ being the
time elapsed under a quench from high-temperature, follows nicely a
slightly-modified power law for $t>t_w$. Very tiny (logarithmic), yet
clearly detectable deviations from the full-aging $t/t_w$ scaling can
be observed. Furthermore, the $t<t_w$ data shows clear indications of
the presence of more than one time-sector in the aging dynamics.
Similar results are found in four-dimensions, but a rather different
behaviour is obtained in the infinite-dimensional $z=6$ Viana-Bray
model.  Most surprisingly, our results in infinite dimensions seem
incompatible with dynamical ultrametricity. A detailed study of the
link correlation function is presented, suggesting that its
aging-properties are the same as for the spin correlation-function.
\end{abstract}

\submitto{\JPA}
\maketitle

\section{Introduction}

Spin-glasses~\cite{EXPBOOK,BOOKS,BOOKRSB,YOUNGBOOK} were discovered to
{\em age} even on human time-scales some twenty years
ago~\cite{AGINGDISCOVER}. Aging is nicely demonstrated, for instance,
in measures of the thermoremanent magnetization (see
e.g.~\cite{SITGES96}): in the presence of a magnetic field, cool an
spin-glass from room temperature to the working temperature, $T$,
below its glass-temperature; hold the magnetic field for a while (the
time elapsed will be called $t_w$ hereafter), then switch-off the
field and record the time-decay of the magnetization $M_{t_w}(t)$. Not
only this decay is very slow but, even for the longest $t_w$ tried up
to now, $M_{t_w}(t)$ strongly depends on $t_w$ (the larger $t_w$ is,
the slower decays $M_{t_w}(t)$). It has slowly become clear that the
important information coming-out from experiments in spin-glasses
regards dynamic out-equilibrium effects, such as this one or the more
sophisticated memory and rejuvenation
effects~\cite{MEM-AND-REJ,SITGES96}. Although there has been a burst
of theoretical activity in out-of-equilibrium
dynamics~\cite{REV-DYN,RIEGER,JJ}, it is still not clear\footnote{A
very encouraging experiment~\cite{FDTEXP} measuring the violation
factor of the fluctuation-dissipation theorem \cite{FRANZ,FDT} has
been recently reported.} how out-equilibrium experiments will help us
to choose among the conflicting theoretical views on the nature of
spin-glasses: Replica-Symmetry Breaking~\cite{BOOKRSB} (RSB), the
droplets picture~\cite{DROPLETS}, and the intermediate TNT
picture~\cite{TNT}.  The situation is farther complicated by the fact
that detailed theoretical predictions (to be confronted with
experiments) can be extracted from models only through Monte Carlo
simulations\cite{RIEGER}. It is worth recalling that numerical results
on out-of-equilibrium dynamics have cast some doubts even on the
usefulness of the Edwards-Anderson model to describe physical
spin-glasses~\cite{APE} (see however~\cite{MEMORY} for some reassuring
results).

It is thus clear that one needs to address quantitatively the
time-decay of $M_{t_w}(t)$ or equivalently, given the
Fluctuation-Dissipation Theorem~\cite{REV-DYN}, the time-dependent
correlation function\footnote{Actually, in the aging regime they are
related by a very smooth function~\cite{FRANZ,FDT}.} in the absence of
a magnetic field:
\begin{equation}
C(t,t_w)=\frac{1}{N} \sum_i \langle \sigma_i (t+t_w)\, \sigma_i(t)\rangle
\end{equation}
Now, it seems to be a fact of general validity in out-equilibrium
dynamics \cite{REV-DYN} that $C(t,t_w)$ behaves differently in
different {\em time-sectors}. Loosing generality\footnote{The here
presented formulation cannot describe logarithmic domain-growth, for
instance. See in \cite{REV-DYN} the general framework.} for the sake
of clarity, this amounts to say that it can be decomposed as
\begin{equation}
C(t,t_w)=\sum_i
f_i\left(\frac{(t+t_w)^{1-\mu_i}-t_w^{1-\mu_i}}{1-\mu}\right)\,.
\end{equation}
Here, $f_i$ are smooth, decreasing functions that tend to zero at
infinity, and such that $f_i(0)$ is of order one. It follows that if
$t\ll t_w^{\mu_i}$ the $i-th$ time sector contributes the constant
value $f_i(0)$ to $C(t,t_w)$, while for $t\gg t_w^{\mu_i}$ it
contributes nothing. In other words, the $i-th$ time sector is active
only for $t\sim t_w^{\mu_i}$ (notice that the different time sectors
get neatly separated only in the limit of very large $t_w$). Not much
is known about the exponents $\mu_i$ defining the different time
sectors. With this popular parametrization~\cite{REV-DYN,SITGES96},
one has $0\le\mu_i\le 1$.  For the simple case of the
coarsening-dynamics (domain-growth) of an ordered
ferromagnet\cite{BRAY}, only two time sectors are needed for a
complete description: $\mu_1=0$ describing the stationary, $t_w$
independent dynamics found at small $t$, and $\mu_2=1$ describing the
{\em full-aging} situation where the correlation-function depends on
the ratio $t/t_w$. Also the spin-glass dynamics has been
experimentally claimed\cite{SITGES96} to be ruled only by two time
sectors: $\mu_1=0$ and $\mu_2=0.97$. The second time sector is
slightly but clearly different from the full-aging $t/t_w$ behavior
($\mu=1$) and is thus named sub-aging. However, a very recent
experiment\cite{RODRIGUEZ} seems to indicate that the sub-aging
behavior is just an artifact of the finite-time needed to cool the
system down to the working temperature (a limitation not suffered of
in numerical simulations). Using their fastest cooling protocol,
Rodriguez et al.\cite{RODRIGUEZ} have found a clear full-aging
behavior $\mu_2=0.999$. Furthermore, the role of the stationary
time-sector ($\mu_1=0$) to describe the data is far less critical than
previously\cite{SITGES96} thought. It is also worth mentioning the
recent numerical results in 3 and 4 dimensions by Berthier and
Bouchaud~\cite{MEMORY}, who found superaging for infinite cooling
rates, turning to subaging for finite cooling rates.  At this point,
we wish to make two comments:
\begin{itemize}
\item The presence of more than two time-sectors seems to be a crucial
requirement for the  validity of the dynamic version of the usual ultrametric Replica-Symmetry 
Breaking description
of spin-glasses with an infinite number of replica symmetry breakings.
\item If the largest $\mu$ exponent is 1, as implied by this popular
parametrization \cite{SITGES96,REV-DYN}, for the very large $t_w$
achieved in experiments (1 second means roughly $t_w=10^{12}$), it is
quite possible that all the faster sectors have already died-out since
$t_w^{\mu_i}\ll t_w\sim t$. A full-aging ansatz could pretty well
describe the data, specially if the second-largest $\mu$ exponent is
significantly smaller than one. In this respect, a numerical
simulation could have a better chance of observing the different time
sectors.
\end{itemize}

To conclude this introduction, let us recall the main results of
previous intensive numerical studies of aging dynamics~\cite{RIEGER}
(for a more recent extensive numerical study of the aging dynamics
with a different focus see~\cite{APE}).  Indeed, it was
found~\cite{RIEGER} that the correlation function at short times
($t\ll t_w$) could be described as a power law with a temperature
dependent exponent ($C(t,t_w)\sim t^{-x(T)}$). The exponent $x(T)$ was
found to be fairly small ($x(T=0.5T_\mathrm{c})=0.015$) On the other
hand, at long times a power-law decay with a different exponent,
$\lambda(T,t_w)$, was observed ($C(t,t_w)=t^{-\lambda(T,t_w)}$ for
$t\gg t_w$). Yet a cross-over functional form was
proposed~\cite{RIEGERFIRST} (see also Ref.~\cite{RIEGER}), implying
that $\lambda(T,t_w)$ saturates to $\lambda(T)$ for large $t_w$:
\begin{equation}
C(t,t_w)= t^{-x(T)} \Phi\left(\frac{t}{t_w}\right)\,.
\label{EQRIEGER}
\end{equation}
The cross-over function was proposed to be decreasing, smooth, and to
have the asymptotic behaviors $\Phi(y)\sim constant$ for small $y$
and $\Phi(y)\sim y^{-\lambda(T)+x(T)}$ for large $y$. However, in the
spanned time scales~\cite{RIEGER} ($t_w \leq 10^5$, $t\leq 10^6$),
$\lambda(T)$ was actually found to significantly depend on $t_w$. Yet,
if one assumes that $\lambda(T,t_w)$ saturates to $\lambda(T)$ for
large $t_w$, so that Eq.(\ref{EQRIEGER}) could hold, one would speak
of a dynamics with effectively three time sectors: $\mu_1=0$,
$\mu_2=1-x(T)/\lambda(T)$, and $\mu_3=1$ (to derive this, assume that
$\Phi(y)$ is an analytical function, such as
$(1+y)^{-\lambda(T)+x(T)}$).  However, since $x(T)/\lambda(T)$ was
found~\cite{RIEGER} to be $0.1$ or smaller, it would not be easy to
separate the $\mu_2$ and $\mu_3$ sectors. On the other hand, it is
clear that Eq.(\ref{EQRIEGER}) describes a non-ultrametric dynamics.

In this paper, by extensive Monte Carlo simulations using among others
the dedicated SUE machine~\cite{CLU,RUCLU,SUE}, we will show that
there are small yet quite significant deviations to
Eq.(\ref{EQRIEGER}) in finite dimensions. Surprisingly enough, our
results in infinite dimensions are rather different, and suggest that
dynamical ultrametricity could {\em not} hold in (some) statically
ultrametric systems. As the reader will notice, heavy use of data
fitting will be made in the following. Unfortunately there is no
precise theory that tell us which should be the precise functional
form of $C(t+t_w, t_w)$, so it is difficult to justify theoretically
many of the fits. Also the very good power like behaviour we found
cannot be recovered from analytic computations. Here we are doing some
kind of exploratory work, trying to guess which could be a reasonable
form that well represents the data. This may be useful for a number of
reasons:  If we can do good fits at given vale of $t_w$ it is rather
more convenient to look to the dependence of the fits parameter on
$t_w$ than to the data themselves (sometimes scaling plots may be
misleading), good fits may evidentiate some behaviour that could be
eventually analytically derived, fits can also used to extrapolate
data to large times. Finally, notice that as an outcome of this
strategy, some of the fits that we are doing could also be useful in
analyzing experimental data.

The layout of the rest of this paper is as follows. In
section~\ref{SIMULSECT} we describe our simulations. In
section~\ref{SECTC} we concentrate on the spin-spin correlation
function, presenting a new parametrization of the function $C(t,t_w)$,
and discussing the possibility of numerically studying the existence of
more than two time-sectors. In section~\ref{SECTQ} we focus on the
aging behavior of the link-overlap and the link-correlation function
(defined in section~\ref{SIMULSECT}). We shall conclude that even in
the limit of infinite waiting time, the link-correlation function
ages.  Finally, we present our conclusions in
section~\ref{SECTCONCLUSIONS}.

\section{The simulation}\label{SIMULSECT}

We have studied the three dimensional Ising spin glass defined on a
cubic lattice ($L\times L \times L$) with helicoidal boundary
conditions\footnote{Let $(x,y,z)$ be the lattice coordinates of spin
number $i\equiv x + y L + z L^2$, then the coordinates of the three
nearest neighbors (in the three positive directions) are given by:
$i_1=(i+1) \,\mathrm{mod}\, L^3 $, $i_2= (i+L) \,\mathrm{mod}\, L^3$
and $i_3= (i+L^2)\, \mathrm{mod}\, L^3$.}.  The Hamiltonian is
\begin{equation}
{\cal H}=-\sum_{<i,j>} \sigma_i J_{ij} \sigma_{j} \ .
\end{equation}
The large time scale and lattice sizes simulated have been possible
due to the use of a dedicated computer, SUE (Spin Update Engine,
Universidad de Zaragoza) based on programmable components, and
achieving an update speed of of 0.22 nanoseconds per spin. Details
about the machine can be found in Refs.~\cite{CLU,RUCLU,SUE}.

The volume of the system is $V=L^3$, $\sigma_i$ are Ising variables,
$J_{ij}$ (uncorrelated quenched disorder) are $\pm 1$ with equal
probability, and the sum is extended to all pairs of nearest
neighbors.  The choice of helicoidal boundary conditions is mandatory
(for us) because the hardware of the SUE machine has been optimized
for them. During the simulation we have measured the following
quantities:
\begin{eqnarray}
C(t,t_w)&=&\frac{1}{N} \sum_i\  \langle \sigma_i (t+t_w)\, \sigma_i(t)
\rangle\,,\\
C_\mathrm{link}(t,t_w)&=&\frac{1}{D N}\sum_{\mu=1}^D \sum_i\  \langle
\sigma_i(t+t_w) \sigma_{i+\mu}(t+t_w)\, \sigma_i(t_w)
\sigma_{i+\mu}(t_w) \rangle\,,\label{EQCLINK} \\
q(t_w)&=&\left|\frac{1}{N} \sum_i\  \langle \sigma_i^{(1)} (t_w) \sigma_i^{(2)}
(t_w)\rangle\,\right|\,,\label{EQQ}\\
 q_\mathrm{link}(t_w)&=& \frac{1}{D N}\sum_{\mu=1}^D \sum_i\ \langle
\sigma_i^{(1)}(t_w) \sigma_{i+\mu}^{(1)}(t_w)\, \sigma_i^{(2)}(t_w)
\sigma_{i+\mu}^{(2)}(t_w) \rangle\,.\label{EQQLINK}
\end{eqnarray}
in the above equations, $D$ is the spatial dimension, $i+\mu$ stands
for the nearest neighbor in the $\mu$ direction, while the superscript
$(1)$ and $(2)$ refer to the replica index (real replicas: pair of
systems evolving independently with the same couplings $J_{i,j}$). The
notation $\langle\ldots\rangle$ refers both to average over thermal
histories (random-numbers) and over disorder realizations. Let us just
recall that since the starting configurations are random, we have
explored the so-called $q=0$ sector~\cite{BOOKRSB}, in which one
expects dynamical correlation functions to be self-averaging.

Given the unique features of the SUE machine we have preferred to use
it for very long runs in a rather small number of samples. The lattice
sizes studied have been $ L=20,L=30$ and $L=60$.  We have considered
three values of the inverse temperature: $\beta=1.25,1.67$ and $2.0$
(hereafter to be referred to as $T=0.8,0.6$ and $T=0.5$
respectively). The critical temperature for this model is
$T_\mathrm{c}=1.14(1)$, thus the selected temperatures are $0.7
T_\mathrm{c}, 0.53 T_\mathrm{c}$ and $0.44 T_\mathrm{c}$ respectively.
Given the very slow growth of the spin-glass coherence length (see
e.g.~\cite{RIEGER,LENGTH}) one should not expect noticeable
finite-size effects even for the $L=20$ lattice. However, as it is
well known, sample-to-sample fluctuations in $C(t,t_w)$ decrease
fastly with growing system sizes.

We can sum up the details of the SUE simulations in table
\ref{SIM_TABLE}.  The SUE time-step corresponds to 8192 full-lattice
sequential heat-bath updates.  For $C(t,t_w)$ and
$C_\mathrm{link}(t,t_w)$ we have selected the $t$ and $t_w$ values in
a logarithmic scale. These values corresponds actually to $8192\times
2^{0.25 i},i=1,\dots$.  The number of simulated samples has been 16
for the $L=60$ systems and 32 for the $L=20,30$ ones. Here, we will
only present the results for $L=60$, since the data for $L=20$ and
$L=30$ are fully compatible with them, but far noisier. Yet, much more
accurate data have been obtained for moderate $t$ and $t_w$ from
simulations on a PC (see below) of an $L=33$ system.

\begin{table}[ht!]
\begin{tabular}{|c|c|c|c|}
\hline
L  &  T & Number of iterations & Number of samples\\ 
\hline
$60$ & $0.5$ & $10^9$ & $16$ \\
\hline
$60$ & $0.6$ & $8\times 10^8$ & $16$ \\
\hline
$60$ & $0.8$ & $5\times 10^8$ & $16$ \\
\hline
$30$ & $0.5$ & $5\times 10^8$ & $32$ \\
\hline
$30$ & $0.8$ & $5.5\times 10^8$ & $32$ \\
\hline
$20$ & $0.5$ & $2\times 10^9$ & $32$ \\
\hline
\end{tabular}
\caption{SUE simulations details}
\label{SIM_TABLE}
\end{table}
To have some data at times shorter than SUE step $8192$, we have
simulated (using heat-bath) on a personal computer 80 samples with  $L=60$
($t_w\leq 6877$ , $t\leq 65519$).

In addition, we have performed Metropolis simulations on personal
computers for the same model in 3D ($L=33$, $T=0.8=0.7 T_\mathrm{c}$,
$t_w\leq 2^{18}$, $t\leq 2^{22}$), 4D ($L=37$, $T=1.2=0.6
T_\mathrm{c}$, $t_w\leq 2^{18}$, $t\leq 2^{22}$ ) and in the
infinite-dimensional Viana-Bray model ($z=6$, $N=5\times 10^6$,
$T=0.8=0.34 T_\mathrm{c}$, $t_w\leq 2^{13}$, $t\leq 2^{17}$ and  $N=10^6$
$T=0.8$, $t_w\leq 2^{19}$, $t\leq 2^{16}$). Even if
the times were shorter than in SUE, the number of simulated samples
has been much larger. The typical statistical error in correlation
functions (as calculated from the sample-to-sample fluctuations)
obtained in the PC was 20 times smaller than the statistical errors of
the SUE results.

\section{Aging dynamics}\label{SECTC}

\begin{figure}[t!]
\begin{center}
\leavevmode
\epsfig{file=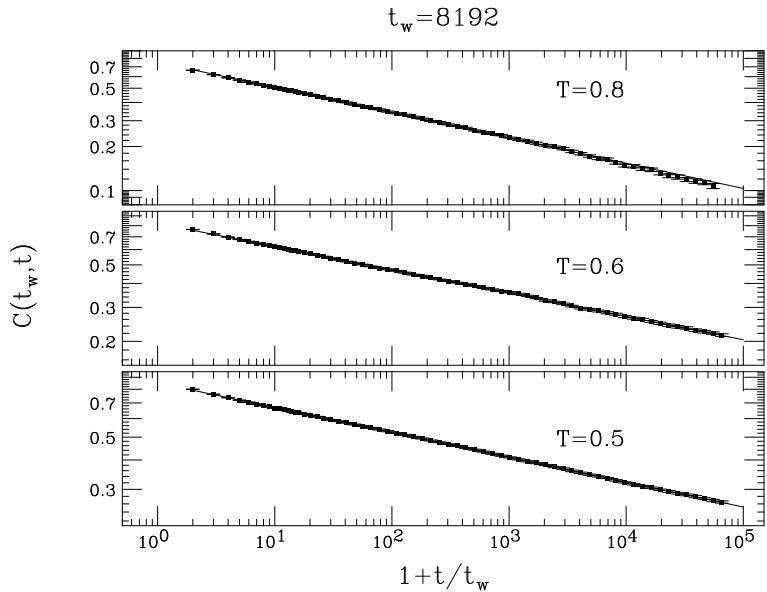,width=0.5\linewidth,angle=90}
\end{center}
\caption{Double logarithmic plot of the correlation-function obtained with
SUE using a heat-bath algorithm, versus $1+t/t_w$, for $t_w=8192$ and
$T=0.5,0.6$ and $0.8$. A single power law fits satisfactorily the data
for $t>t_w$ at all-three temperatures.}
\label{FIG1}
\end{figure}

\begin{figure}[t!]
\begin{center}
\leavevmode
\epsfig{file=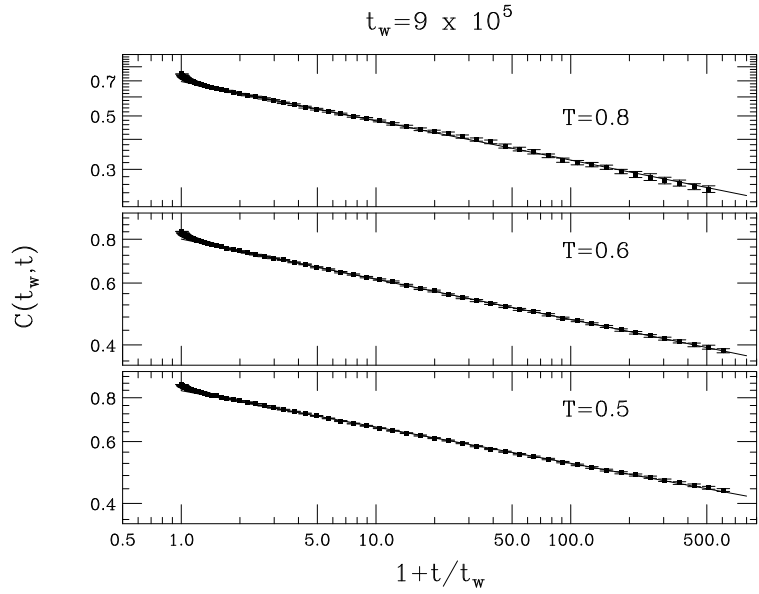,width=0.5\linewidth,angle=90}
\end{center}
\caption{As in Fig.\protect{\ref{FIG2}}, for  $t_w=884736$.}
\label{FIG2}
\end{figure}

\begin{figure}[t!]
\begin{center}
\leavevmode
\epsfig{file=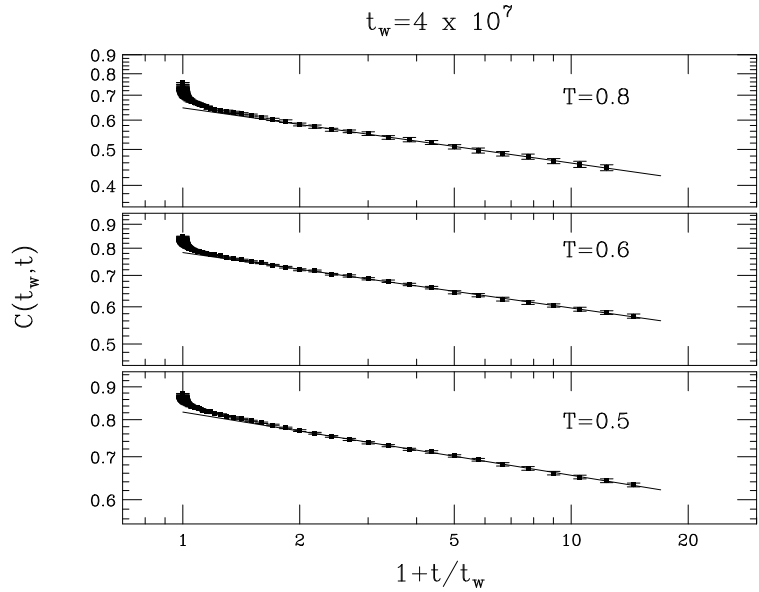,width=0.5\linewidth,angle=90}
\end{center}
\caption{As in Fig.\protect{\ref{FIG2}}, for $t_w=39813120$.}
\label{FIG3}
\end{figure}

We have found in finite dimensions ($D=3$ and $D=4$) that the
correlation function $C(t,t_w)$ can be nicely fitted for $t\geq t_w$
as
\begin{equation}
C(t,t_w)=A(t_w) \left(1+\frac{t}{t_w}\right)^{-1/\alpha(t_w)}\,,\ t>t_w.
\label{LAECUACION}
\end{equation}
As it can be seen in Figs.~\ref{FIG1},\ref{FIG2} and \ref{FIG3}, for a
wide range of $t/t_w$ and $t_w$ and all three temperature a simple
power-law decay seems enough to describe the data, although the
coefficients $A(t_w)$ and $\alpha(t_w)$ clearly depends on temperature
(notice that $\alpha(t_w)$ is just the inverse of Rieger et
al.~\cite{RIEGER} $\lambda(T,t_w)$ exponent). The behavior in
infinite dimensions ($z=6$ Viana-Bray model) is rather different and
will be discussed at the end of this section.

\begin{figure}[t!]
\begin{center}
\leavevmode
\epsfig{file=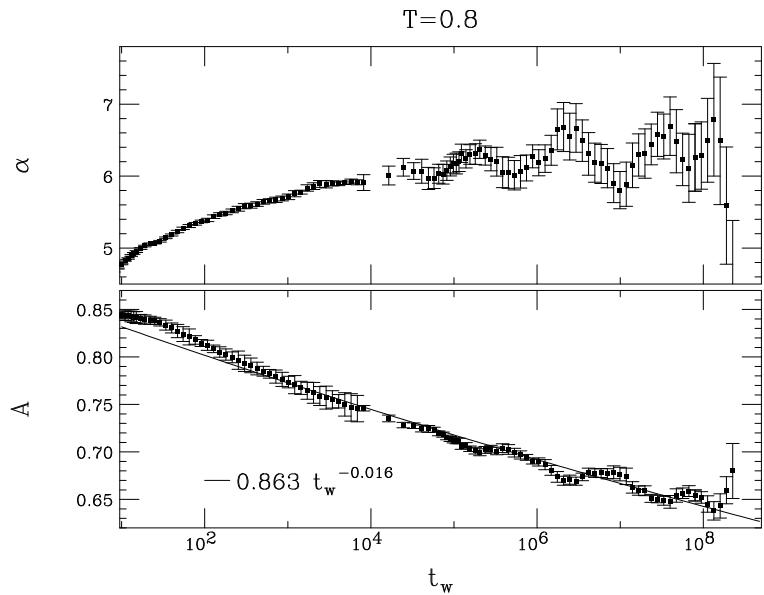,width=0.5\linewidth,angle=90}
\end{center}
\caption{Top: exponent $\alpha(t_w)$ defined in
Eq.(\protect{\ref{LAECUACION}}) versus $t_w$ for the heat-bath
dynamics of the Edwards-Anderson model in $D=3$, at
$0.7 T_\mathrm{c}$. Data for $t_w<8192$ were obtained on a PC, while
for larger $t_w$ correlation-functions from SUE were used. Bottom:
same as in top panel for the prefactor $A(t_w)$, defined in
Eq.(\protect{\ref{LAECUACION}}). The solid line was obtained
fitting for $t_w\geq 8192$.
}
\label{FIG4}
\end{figure}

It is clear that the waiting-time dependence of the prefactor $A(t_w)$
and the exponent $\alpha(t_w)$ is of utmost importance. Should
$A(t_w)$ and $\alpha(t_w)$ tend to constant non vanishing values,
dynamic ultrametricity~\cite{REV-DYN} would not hold, implying that
the usual dynamic formulation of the ultrametric approach to
continuous Replica Symmetry Breaking should be modified. Furthermore,
we have been unable of finding a divergence law for $\alpha(t_w)$
compatible with dynamic ultrametricity, if $A(t_w)$ has a
non-vanishing limiting value for large $t_w$. Thus, we tend to believe
that dynamical ultrametricity implies that $A(t_w)$ should vanish in
the large $t_w$ limit.

To obtain the coefficients $A(t_w)$ and $\alpha(t_w)$ we have fitted
the correlation functions $C(t,t_w)$ to the functional form in
Eq.(\ref{LAECUACION}). Yet, although Figs. ~\ref{FIG1},\ref{FIG2} and
\ref{FIG3}, suggest a pure power-law behavior, we have found some
dependence of $\alpha(t_w)$ on the fitting-window, particularly for
$T=0.5$. Specifically, at $T=0.5$ $\alpha(t_w=8192)$ grows a $10\%$ if
the fit is performed for $t_w\leq t\leq 10^5 t_w$ as compared to the
fit in the window $t_w\leq t\leq 10t_w$. Not taking care of this could
be dangerous since, obviously, the larger $t_w$ the shorter the
achievable $t/t_w$.  Thus, in order to isolate the $t_w$-dependence we
have restricted ourselves to the fitting range $t_w\leq t\leq 10t_w$.
On the other hand, the prefactor $A(t_w)$ is basically independent of
the fitting-window.  Another tricky point is the error-estimate for
$A(t_w)$ and $\alpha(t_w)$.  It is clear that standard techniques
($\chi^2$ minimization) work poorly for tremendously correlated
stochastic variables such as $C(t,t_w)$ for successive $t$. In order to
have a (hopefully) reasonable estimate we have turned to a Jack-Knife
procedure on the fitted coefficients themselves.

\begin{figure}[t!]
\begin{center}
\leavevmode
\epsfig{file=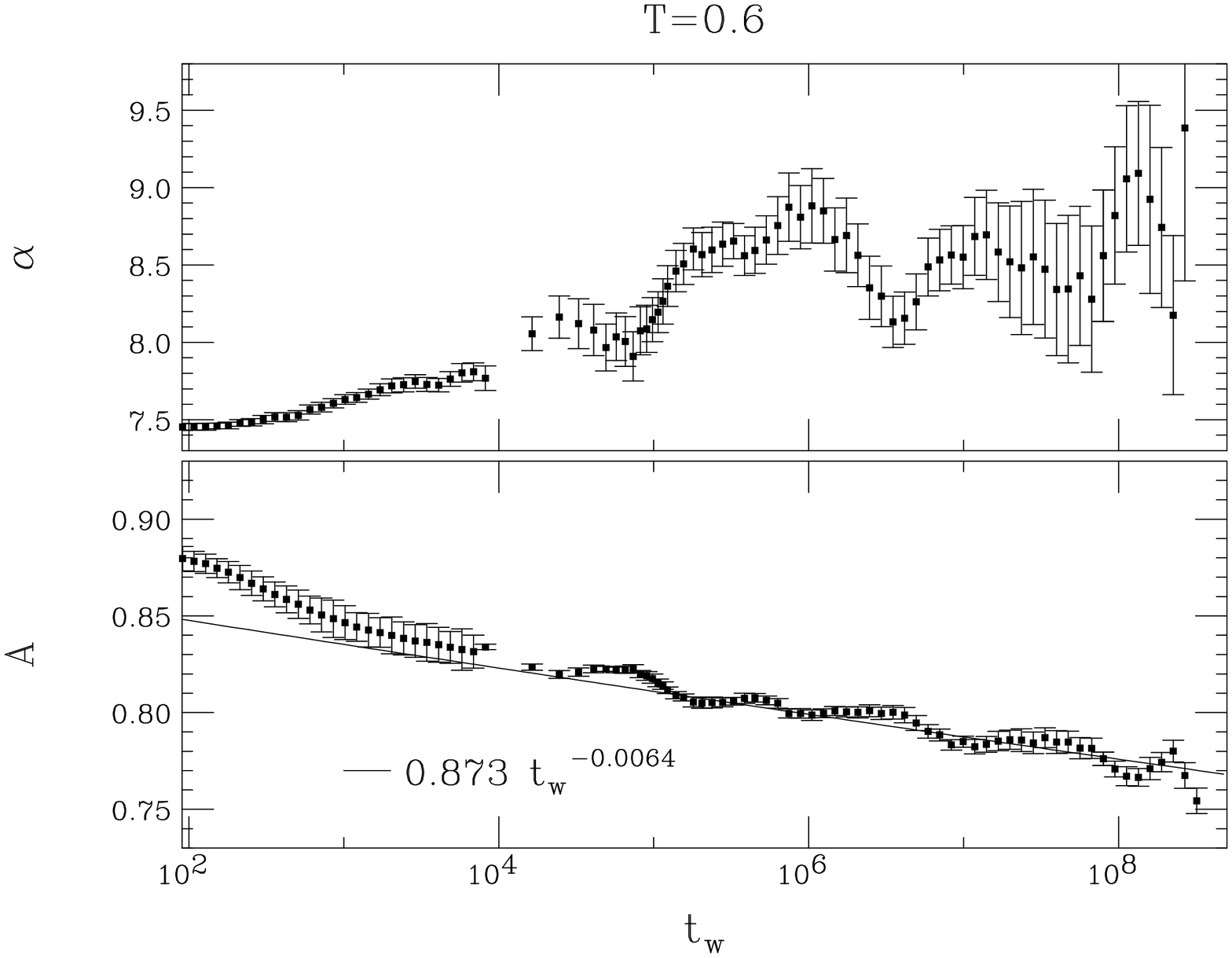,width=0.5\linewidth,angle=90}
\end{center}
\caption{As in Fig.\protect{\ref{FIG4}} for
$0.53 T_\mathrm{c}$.
}
\label{FIG5}
\end{figure}

\begin{figure}[t!]
\begin{center}
\leavevmode
\epsfig{file=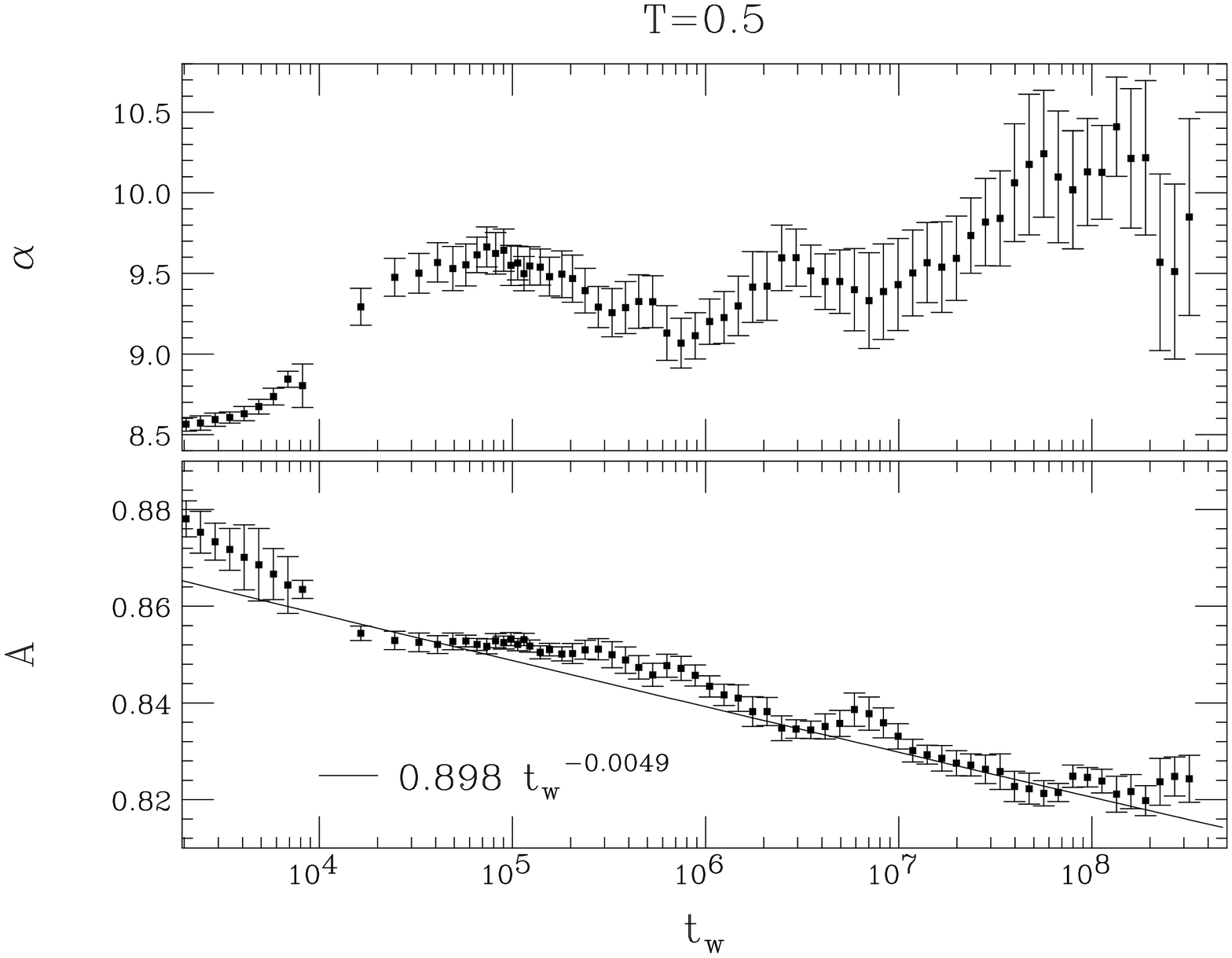,width=0.5\linewidth,angle=90}
\end{center}
\caption{As in Fig.\protect{\ref{FIG4}} for
$0.44 T_\mathrm{c}$.
}
\label{FIG6}
\end{figure}

Our results for the prefactor $A(t_w)$ and the exponent $\alpha(t_w)$
of the heat-bath dynamics in the $D=3$ case, can be found in
Figs.~\ref{FIG4} ($T=0.8$),~\ref{FIG5} ($T=0.6$) and ~\ref{FIG6}
($T=0.5$). The results for $t_w<8192$ were obtained on a PC (see
section~\ref{SIMULSECT}), while for $t_w\geq 8192$ SUE data were used.
Actually, we have found that the functional form (\ref{LAECUACION}) is
not suitable for very small $t_w$, thus there is a lower,
temperature-dependent, cut-off on the $t_w$ values shown in
Figs.~\ref{FIG4},~\ref{FIG5} and~\ref{FIG6}.

As far as the exponent $\alpha(t_w)$ is concerned (see the upper part
of Figs.~\ref{FIG4},~\ref{FIG5} and~\ref{FIG6}), we have a significant
growth for $t_w< 10^6$. For larger $t_w$, $\alpha(t_w)$ is constant
within errors. This (asymptotic?) value would correspond to the
inverse of the large waiting time limit of Rieger et al.~\cite{RIEGER}
aging exponent $\lambda(T,t_w)$. Interestingly enough, the values of
$\alpha(t_w)$ for the largest achieved $t_w$ seems to be proportional
to $T_\mathrm{c}/T$.  On the other hand, the prefactor $A(t_w)$ (see
the lower part of Figs.~\ref{FIG4},~\ref{FIG5} and~\ref{FIG6}) clearly
decreases in all the simulated $t_w$ range. A power law seems to be
appropriate for this decay, the exponent being precisely the exponent
$x(T)$ in Eq.(\ref{EQRIEGER}). Thus, we find $x(0.7\,
T_\mathrm{c})=0.016$, $x(0.53\, T_\mathrm{c})=0.0064$,
$x(0.44\,T_\mathrm{c})=0.0049$.

\begin{figure}[t!]
\begin{center}
\leavevmode
\epsfig{file=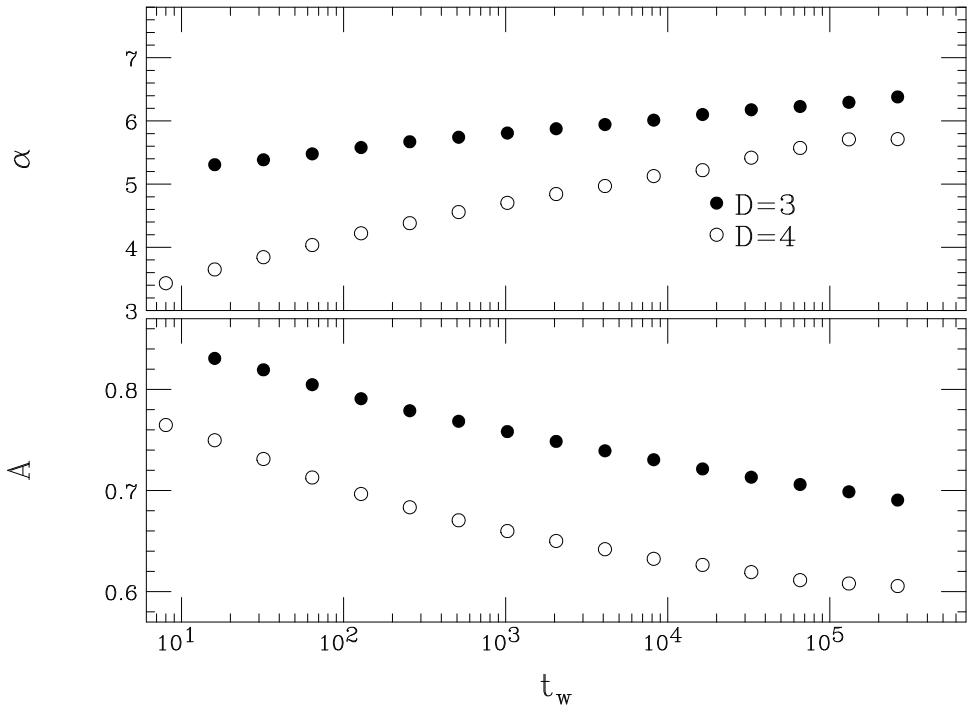,width=0.6\linewidth,angle=90}
\end{center}
\caption{Top: exponent $\alpha(t_w)$ defined in
Eq.(\protect{\ref{LAECUACION}}) versus $t_w$ for the Metropolis
dynamics of the Edwards-Anderson model in $D=3$ at $T=0.8$ (full
circles) and $D=4$ at $T=1.2$ (empty circles). Bottom: same as in top panel for the
prefactor $A(t_w)$, defined in Eq.(\protect{\ref{LAECUACION}}).} 
\label{FIG7}
\end{figure}

We have found similar results (see Fig.~\ref{FIG8}) for the Metropolis
dynamics of the Edwards-Anderson model with binary-couplings
distributions in three dimensions ($T=0.7 T_\mathrm{c}$),
four-dimensions ($T=0.6 T_\mathrm{c}$). For all the simulated $t_w$
($<10^6$) we have found a growing trend in the exponent
$\alpha(t_w)$. As usual, the dynamics in four dimensions is faster
than in three dimensions (notice how the $\alpha(t_w)$ is smaller but
more rapidly growing). The prefactor $A(t_w)$ decreases.  A power-law
seems to be adequate for this decay in $D=3$, but this is less clear
for the four-dimensional lattice.

\begin{figure}[t!]
\begin{center}
\leavevmode
\epsfig{file=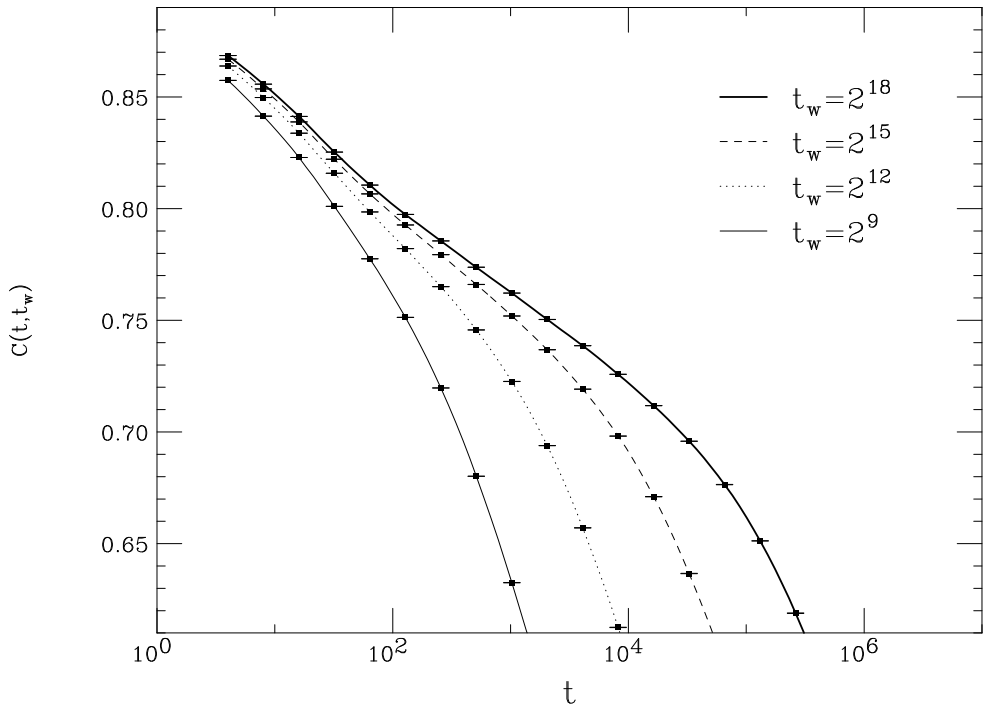,width=0.6\linewidth,angle=90}
\end{center}
\caption{Top: The spin correlation function as a function of $t$ for
several $t_w$, obtained with the Metropolis dynamics of the 3D
Edwards-Anderson model at $T=0.7 T_\mathrm{c}$ (the lines are mere
guide-to-eye). Data were obtained in a $L=33$ lattice simulated on a
PC. The plot concentrates on the region of large $C$ in order to
discuss the time-traslation invariant regime (see text).}
\label{FIGPREPARATION}
\end{figure}

\begin{figure}[t!]
\begin{center}
\leavevmode
\epsfig{file=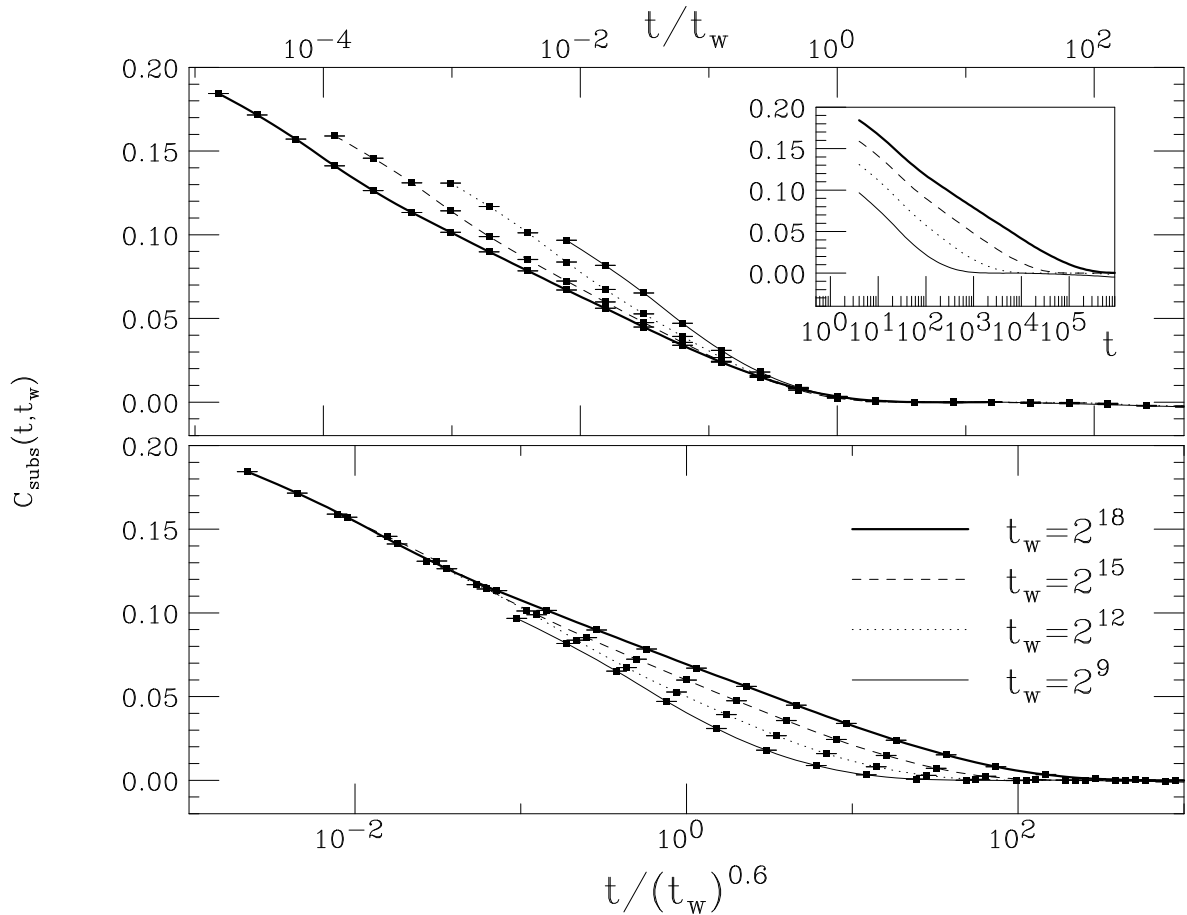,width=0.78\linewidth,angle=90}
\end{center}
\caption{Top: The subtracted correlation function defined in
Eq.(\protect\ref{EQCSUBS}) as a function of $t/t_w$, for the
Metropolis dynamics of the 3D Edwards-Anderson model at $T=0.7
T_\mathrm{c}$ (the lines are mere guide-to-eye). The inset shows
$C_\mathrm{subs}(t,t_w)$ as a function of time, for the different
$t_w$. Bottom: $C_\mathrm{subs}(t,t_w)$, from the same simulation as above,
versus the (not dimensionless!) ratio $t/t_w^{0.6}$.}
\label{FIG8}
\end{figure}

Up to now, we have only confirmed that the dynamics for $t\geq t_w$,
and $t_w > 10^6$ can be rather well described by the
cross-over~\cite{RIEGERFIRST,RIEGER} formula (\ref{EQRIEGER}), and we have given
an explicit form for the cross-over function $\Phi(y)$ (see
Eq.(\ref{LAECUACION})). We are thus predicting that $C(t_w,t_w)$
should vanish in the infinite waiting-time limit: dynamic
ultrametricity reduces to a trivial $0=0$ statement in the $t\geq t_w$
time-sector. Therefore, non trivial statements about dynamical
ultrametricity should regard times $1\ll t\ll t_w$. In this time
sector, the cross-over formula (\ref{EQRIEGER}), predicts a
time-translational invariant ($\mu=0$) power law decay,
$C(t,t_w)=\Phi(0)/t^{x(T)}$. This prediction is of course
non-ultrametric.

However, one could wonder about the presence of more than one time
sector. In fact, as one can see in Fig.~\ref{FIGPREPARATION} for the
Metropolis dynamics of the 3D Edwards-Anderson model at $T=0.7
T_\mathrm{c}$, small but measureable deviations from time-traslation
invariance appear at $t/t_w \sim 10^{-3}$. In order to explore
the regime $1\ll t\ll t_w$, we have introduced a subtracted
correlation function:
\begin{equation}
C_\mathrm{subs}(t,t_w)= C(t,t_w)- A(t_w) \left(1+\frac{t}{t_w}\right)^{1/\alpha(t_w)}\,.
\label{EQCSUBS}
\end{equation}
where $A(t_w)$ and $\alpha(t_w)$ are of course the coefficients
obtained in the fit to the functional form (\ref{LAECUACION}) for each
$t_w$.  To motivate it, let us recall that the quasi equilibrium
regime is realized in the case where $t_w$ goes to infinity at fixed $t$.
Naive scaling predicts that $ C(t+t_w, t_w)$ goes to a non trivial
function of $t/t_w$ when $t_w$ goes to infinity.  However we have
already remarked that we could have also a subaging contribution and
find a non trivial behaviour in the region where $t$ is of order
$t_w^\mu$ with $\mu<1$.  In this region we have $1\ll t \ll t_w$, but
we may be not in the quasi-equilibrium regime as far as $t$ goes to
infinity and it is not fixed. One reason for studying
$C_\mathrm{subs}(t,t_w)$ is that in the case of multitime sectors, the
correlation function is the sum of contributions coming from each time
sector and the subtraction help to identify the given time sector.

We show in Fig.~\ref{FIG8} $C_\mathrm{subs}(t,t_w)$ for $D=3$,
$T=0.8$, as a function of $t/t_w$ (upper-part) and as a function of
$t/t_w^{0.6}$ (lower panel). It is clear that $C_\mathrm{subs}$ can be
described as a function of $t/t_w$ for $t>0.1 t_w$, but that strong
deviations are present for smaller times. The inset in the top panel
of Fig.~\ref{FIG8} shows as well that $C_\mathrm{subs}(t,t_w)$ is {\em
not} a function of $t$, as the cross-over formula (\ref{EQRIEGER})
would suggest. On the other hand, the lower part of Fig.~\ref{FIG8}
shows that $C_\mathrm{subs}(t,t_w)$ seems really a
function\footnote{Notice that the numerical value of the quotient
$t/t_w^{0.6}$ is not invariant under a change of time units, so that
one should really speak about $t/(\tau_0^{0.4} t_w^{0.6})$, $\tau_0$
being the used time unit. Thus, we do not attribute any special
significance to $t/t_w^{0.6}=1$.} of $t/t_w^{0.6}$ during two decades
(corresponding to the decay from $C_\mathrm{subs}=0.18$ to
$C_\mathrm{subs}=0.1$). Up to our knowledge, this is the first time
that a time sector different from $\mu=0$ (time-translational
invariant) and $\mu=1$ (full-aging), has been studied.  A cautionary
remark is in order, though, for one cannot exclude that the
$t/t_w^{0.6}$ scaling could stop to apply at much larger $t_w$.
Unfortunately, because of the small number of simulated samples, we
have not been able to repeat this analysis with SUE data: the $t/t_w$
scaling and the $t/t_w^{0.6}$ scaling for $C_\mathrm{subst}$ look
indistinguishable within error bars.

It is clear that more work needs to be done in order to design an
efficient protocol to study the different time-sectors. Yet, we hope
that the reader will be convinced that it should be feasible.

\begin{figure}[t!]
\begin{center}
\leavevmode
\epsfig{file=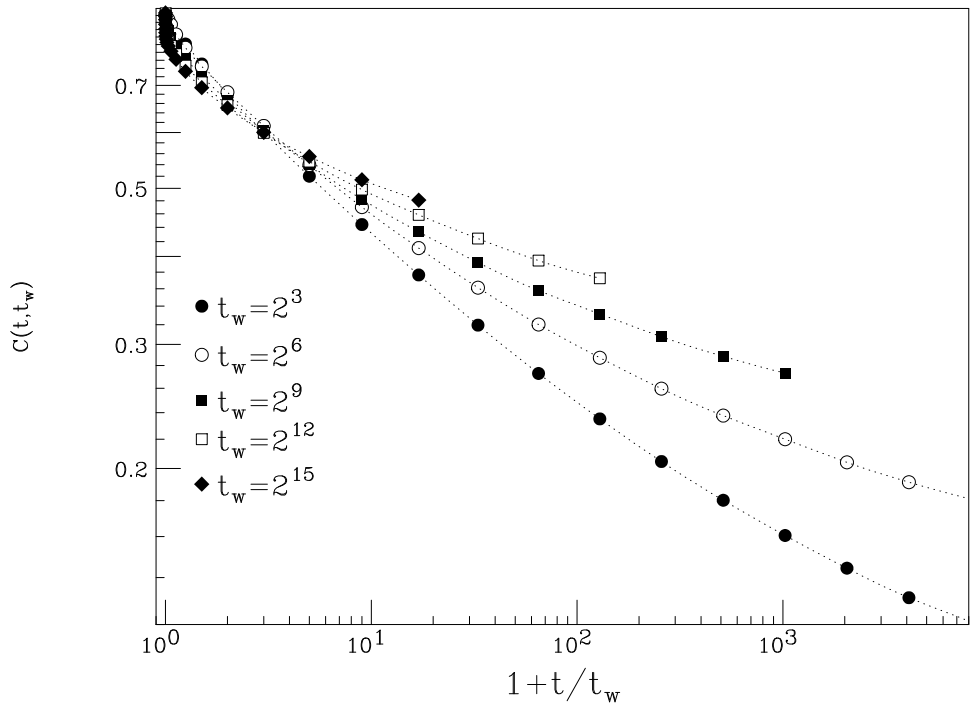,width=0.6\linewidth,angle=90}
\end{center}
\caption{Double logarithmic plot of $C(t,t_w)$ versus $1+t/t_w$ for
the Metropolis dynamics of the infinite-dimensional $z=6$ Viana-Bray
model in ($N=10^6$, $T=0.8=0.34 T_\mathrm{c}$). The clearly measurable
curvature indicates deviations from Eq.(\protect\ref{LAECUACION}).}
\label{FIGVBNO}
\end{figure}

\begin{figure}[t!]
\begin{center}
\leavevmode
\epsfig{file=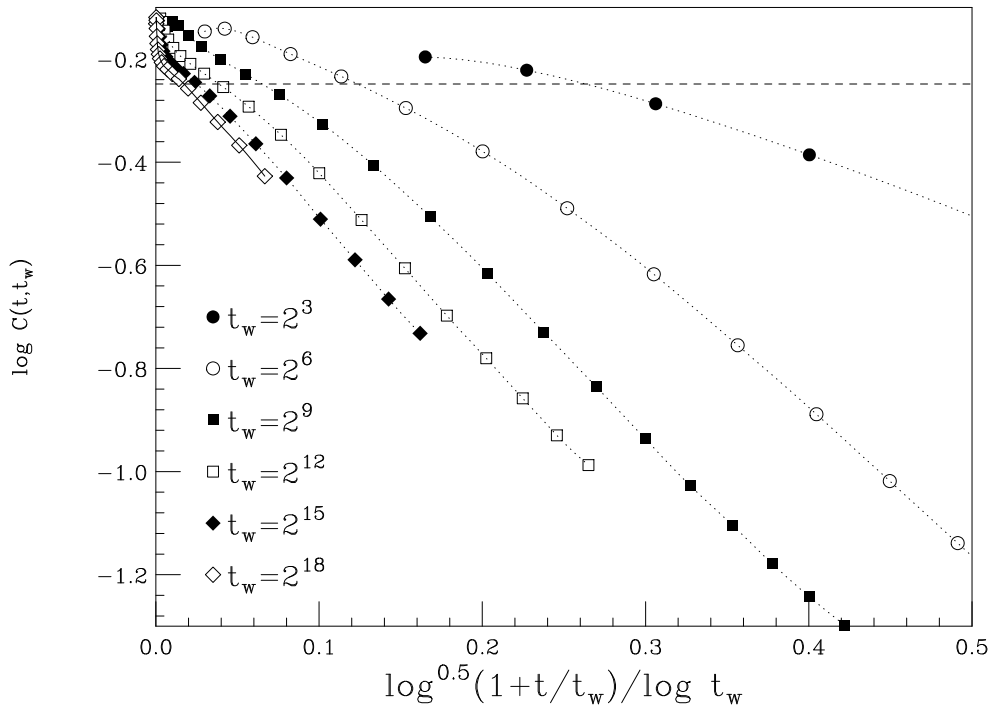,width=0.6\linewidth,angle=90}
\end{center}
\caption{ Logarithm of of $C(t,t_w)$ versus $\sqrt{\log(1+t/t_w)}/\log
t_w$ (see Eq.(\protect\ref{FORMAVB})) for the Metropolis dynamics of
the infinite-dimensional $z=6$ Viana-Bray model in ($N=10^6$,
$T=0.8=0.34 T_\mathrm{c}$). The dashed horizontal line is the logarithm of
the $q_\mathrm{EA}=0.78$~\protect\cite{GIORGIOMARC}.}
\label{FIGVBSI}
\end{figure}

The Metropolis dynamics of the infinite-dimensional $z=6$ Viana-Brey
model is rather different. As we show in fig.~\ref{FIGVBNO}, the
long-time decay of $C(t,t_w)$ is {\em not} a power-law. Thus, it cannot
be fitted with Eq.(\ref{LAECUACION}). We have found (see fig.~\ref{FIGVBSI})
that a fit to
\begin{equation}
C(t,t_w)=S(t_w) \exp\left[-\eta(t_w)\frac{\sqrt{\log\left(1+t/t_w\right)}}{\log t_w}\right]\,,\ t>t_w.
\label{FORMAVB}
\end{equation}
is rather adequate. Moreover, $S(t_w)$ and $\eta(t_w)$ seem to have a
well defined limit for large $t_w$, $S(t_w\to\infty)$ being quite
close to the Edwards-Anderson order-parameter $q_\mathrm{EA}=0.78$
that in this model has been computed~\cite{GIORGIOMARC} at the
one-step level of replica-symmetry breaking. If one believes that
our results are almost asymptotic, so that the large $t_w$ correlation
function truly is
\begin{equation}
C(t,t_w) \approx q_\mathrm{EA}  \exp\left[-\eta\frac{\sqrt{\log\left(1+t/t_w\right)}}{\log t_w}\right]\,,\ t>t_w,
\label{LIMITING}
\end{equation}
the rather amazing conclusion is reached that this ultrametric system
(from the point of view of statics~\cite{BOOKRSB}) is not ultrametric
from the point of view of dynamics! It has been pointed
out~\cite{FRANZ} that dynamical ultrametricity implies (under reasonable
hypothesis) statical ultrametricity, however our results suggest that
dynamical ultrametricity is maybe not a necessary condition for the validity
of RSB. The reader can check that the weaker property of separation of
time scales holds: let $t(C;t_w)$ be the time necessary for the correlation
function at waiting-time  $t_w$ to reach the value $C$. One has
\begin{equation}
\lim_{t_w\to\infty}\frac{t(C_1;t_w)}{t(C_2;t_w)}=0\quad\mathrm{if}\quad
C_1>C_2\,.
\end{equation}
It is somehow disappointing, though, that this property is expected to
hold as well for the Langevin dynamics of the disordered
ferromagnet~\cite{REV-DYN}.

\section{The link overlap}\label{SECTQ}

\begin{figure}[t!]
\begin{center}
\leavevmode
\epsfig{file=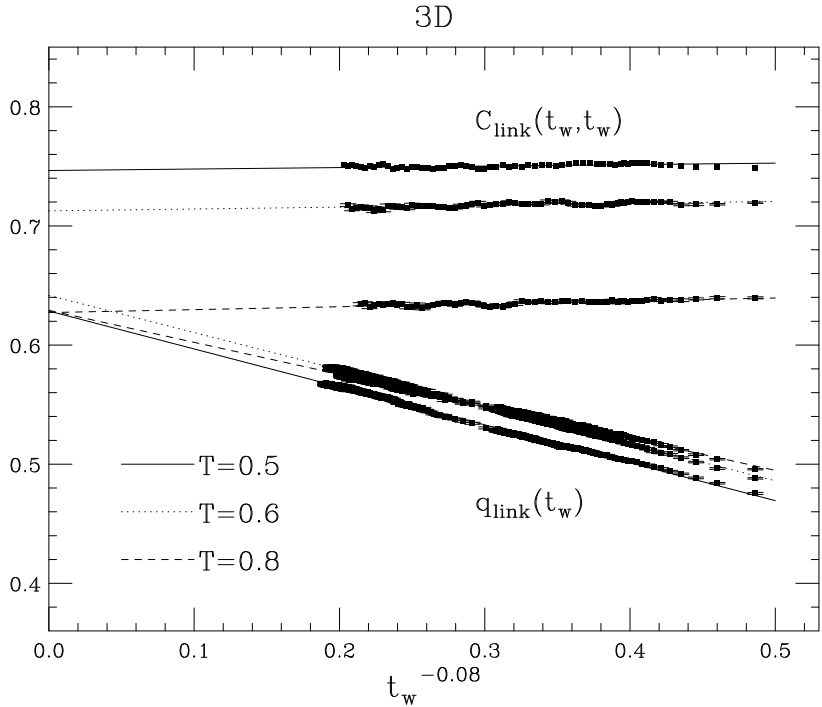,width=0.6\linewidth,angle=90}
\end{center}
\caption{$C_\mathrm{link}(t_w,t_w)$ and $q_\mathrm{link}(t_w)$ versus
$t_w^{-0.08}$, for the heat-bath dynamics of the 3D Edwards-Anderson
model at temperatures $T=0.7 T_\mathrm{c}$, $0.53 T_\mathrm{c}$ and
$0.44 T_\mathrm{c}$. Lines are linear fits.}
\label{FIGEXTRAPOLQC3D}
\end{figure}

\begin{figure}[t!]
\begin{center}
\leavevmode
\epsfig{file=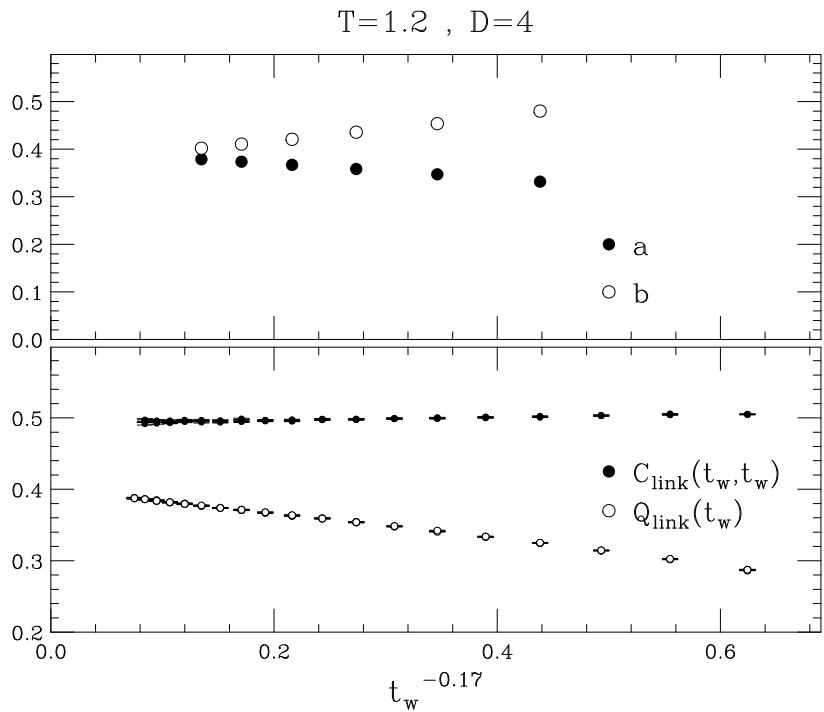,width=0.6\linewidth,angle=90}
\end{center}
\caption{Top: Coefficients $a(t_w)$ and $b(t_w)$ (defined in
Eq.(\protect\ref{COEFICIENTESC2CLINK}), versus $t_w^{-0.17}$, for the
Metropolis dynamics of the 4D Edwards-Anderson model at temperatures
$T=0.6 T_\mathrm{c}$. The fits have been performed in the window $0.1<
C^2(t,t_w)<0.3$.  Bottom: $C_\mathrm{link}(t_w,t_w)$ and
$q_\mathrm{link}(t_w)$ versus $t_w^{-0.17}$, for the Metropolis
dynamics of the 4D Edwards-Anderson model at temperatures $T=0.6
T_\mathrm{c}$.}
\label{FIGCOEF4D}
\end{figure}

\begin{figure}[t!]
\begin{center}
\leavevmode
\epsfig{file=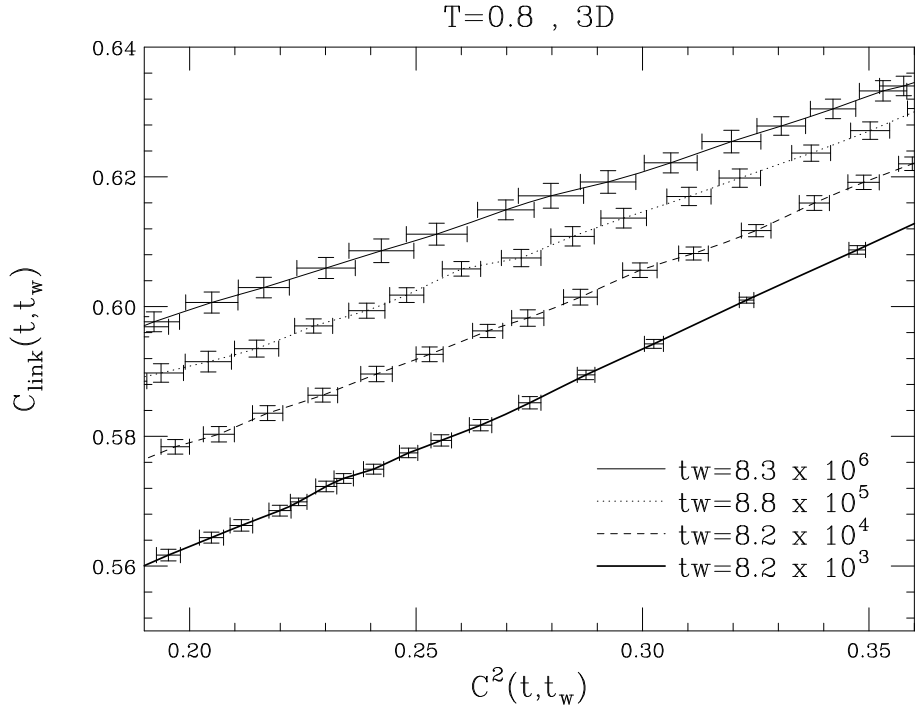,width=0.6\linewidth,angle=90}
\end{center}
\caption{$C_\mathrm{link}(t,t_w)$ versus $C^2(t,t_w)$, obtained in SUE
for the heat-bath dynamics of the 3D Edwards-Anderson model at $T=0.7
T_\mathrm{c}$. The plot is restricted to the window of $C^2$ where a
linear dependence is found for all the studied $t_w$. The lines are
linear fits.}
\label{FIGC2CLINK}
\end{figure}

\begin{figure}[t!]
\begin{center}
\leavevmode
\epsfig{file=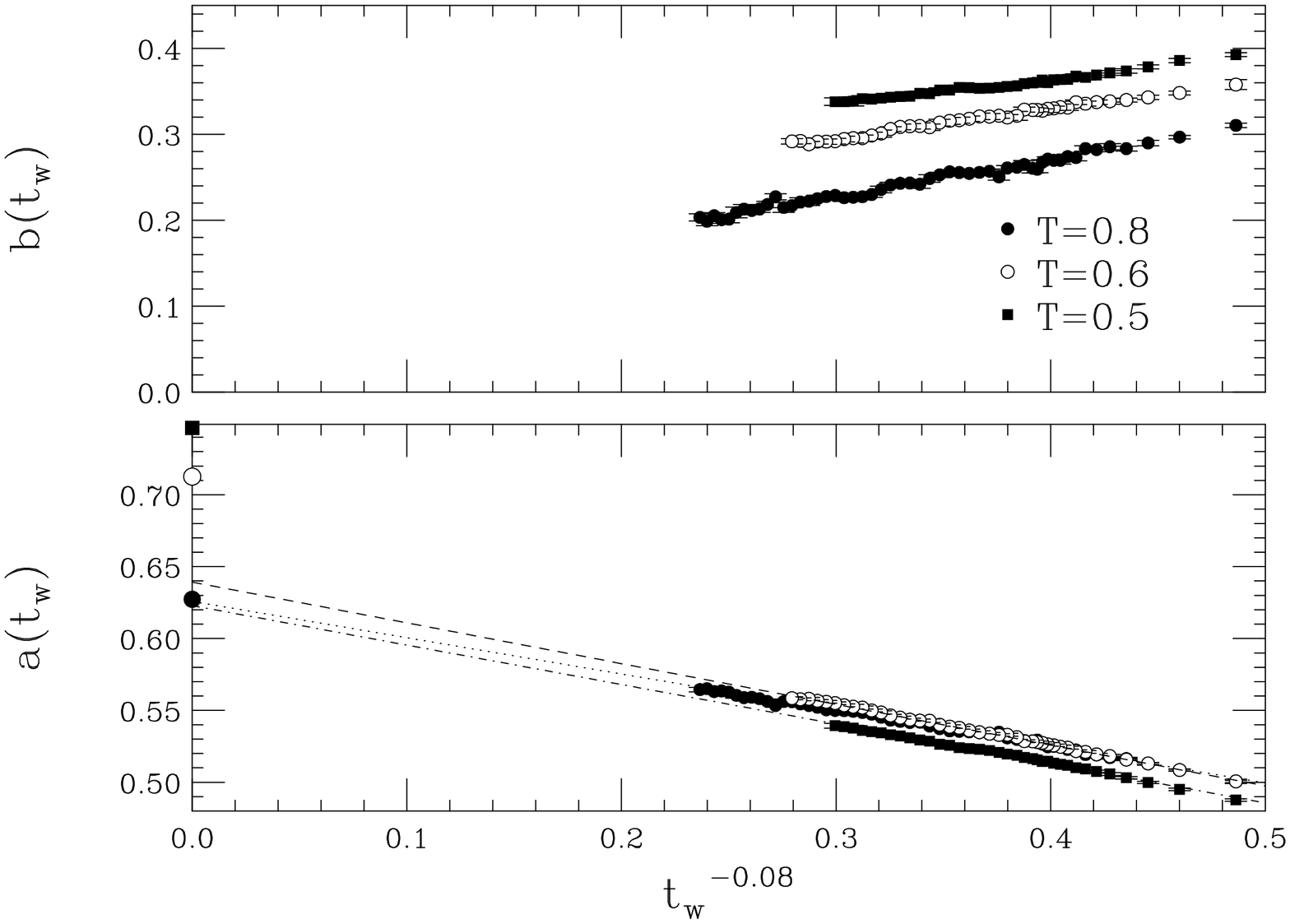,width=0.6\linewidth,angle=90}
\end{center}
\caption{Top: Coefficient $b(t_w)$, defined in
Eq.(\protect\ref{COEFICIENTESC2CLINK}) obtained in SUE for the
heat-bath dynamics of the 3D Edwards-Anderson model versus
$t_w^{-0.08}$, for temperatures $T=0.7 T_\mathrm{c}$, $0.53
T_\mathrm{c}$ and $0.44 T_\mathrm{c}$. Bottom: As in top part, for
coefficient $a(t_w)$, defined in
Eq.(\protect\ref{COEFICIENTESC2CLINK}). The lines are linear fits. We
also plot in the vertical axis the infinite-limit extrapolation for
$C_\mathrm{link}(t_w,t_w)$ (see Fig.~\protect{\ref{FIGEXTRAPOLQC3D}})
for all three temperatures. Notice that for $T=0.8$ the infinite $t_w$
limit of $a(t_w)$ and of $C_\mathrm{link}(t_w,t_w)$ seem compatible.}
\label{FIGCOEF3D}
\end{figure}

In recent years, a new picture of the low temperature phase of
spin-glasses has been put forward~\cite{TNT}, the so-called TNT
picture. These authors propose that the link-overlap defined in
Eq.(\ref{EQQLINK}) should have a trivial distribution function (a
Dirac delta) in the thermodynamic limit (for a recent study of the
link-overlap and related quantities see Ref.~\cite{JJ2002}).  On the
other hand, the spin-overlap defined in Eq.(\ref{EQQ}), would have a
non trivial distribution function, as predicted by RSB~\cite{BOOKRSB}.
In contradiction with RSB, the TNT picture requires that the large
waiting time limit of $C_\mathrm{link}(t_w,t_w)$ be equal to the limit
of $q_\mathrm{link}(t_w)$ (since there would be only a possible value
for this quantity!). This can be checked in our simulations. In
Fig.~\ref{FIGEXTRAPOLQC3D} we show $C_\mathrm{link}(t_w,t_w)$ and
$q_\mathrm{link}(t_w)$ obtained from SUE, for temperatures $T=0.7
T_\mathrm{c}$, $T=0.53 T_\mathrm{c}$ and $T=0.44 T_\mathrm{c}$, as a
function of $t_w^{-0.08}$. The extrapolation as a function of
$t_w^{-0.08}$ was suggested by the fact that $q_\mathrm{link}(t_w)$ is
roughly a linear function of this variable. Some arguments for its
validity has also been given in previous work~\cite{JJ2002,JJa,LENGTH}. We
notice that $q_\mathrm{link}(t_w)$ has a very mild temperature
dependence. Furthermore, for large $t_w$, $q_\mathrm{link}(t_w,T=0.6)$
and $q_\mathrm{link}(t_w,T=0.5)$ are on the top of each other (within
error bars), while $q_\mathrm{link}(t_w,T=0.8)$ is fast approaching
them. Indeed, one would expect $q_\mathrm{link}(t_w=\infty,T=0.5) >
q_\mathrm{link}(t_w=\infty,T=0.6) >
q_\mathrm{link}(t_w=\infty,T=0.8)$, so that the three lines should
cross. The fact that the infinite waiting-time limit of
$q_\mathrm{link}(t_w,T=0.5)$ is smaller than
$q_\mathrm{link}(t_w=\infty,T=0.6) >
q_\mathrm{link}(t_w=\infty,T=0.8)$, indicates that for $T=0.5$ the
data should probably extrapolate like $t_w^{-\eta}$ with
$\eta<0.08$. On the other hand, $C_\mathrm{link}(t_w,t_w)$ shows a
much stronger temperature dependence and, for $T=0.5$ and $T=0.6$, is
basically $t_w$ independent. It seems plausible that, for $T=0.8$, the
limiting value of $C_\mathrm{link}(t_w,t_w)$ and
$q_\mathrm{link}(t_w)$ will be fairly close. However, for $T=0.5$ and
$0.6$ the limits will be noticeably different unless the dynamics
changes drastically at larger $t_w$. Interestingly enough, in four
dimensions and $T=0.6 T_\mathrm{c}$ (see Fig.~\ref{FIGCOEF4D}, lower
part), $C_\mathrm{link}(t_w,t_w)$ and $q_\mathrm{link}(t_w)$ seems to
be linear in $t_w^{-0.17}$, and to extrapolate to different values.

Another interesting question regards the aging properties of the link
correlation-function. From the droplet~\cite{DROPLETS} and TNT
pictures of spin-glasses, one would not expect that
$C_\mathrm{link}(t,t_w)$ would age (at least, for large enough
$t_w$). On the other hand, the RSB picture expects aging properties
akin to the ones of $C(t,t_w)$. To check for this, one could just look
to $C_\mathrm{link}(t,t_w)$ as a function of $C^2(t,t_w)$ (see
Fig.~\ref{FIGC2CLINK}). If one concentrates in a small window of
$C^2$, $0.2 < C^2 < 0.35$ a linear description is perfectly adequate:
\begin{equation}
C_\mathrm{link}(t,t_w)=a(t_w) + b(t_w) C^2(t,t_w)\,.
\label{COEFICIENTESC2CLINK}
\end{equation}
The question now translates to the behavior of $a(t_w)$ and $b(t_w)$.
The TNT and droplet pictures predict that $b(t_w)$ tend to zero and
$a(t_w)$ tend to the large waiting time limit of
$C_\mathrm{link}(t_w,t_w)$.  On the other hand, RSB predicts a
non-vanishing limit of $b(t_w)$.

In Fig.\ref{FIGCOEF3D}, we show coefficients $a(t_w)$ (bottom) and
$b(t_w)$ (top), as a function of $t_w^{-0.08}$, for the heat-bath
dynamics of the 3D Edwards-Anderson model, at $T=0.5, 0.6$ and $0.8$.
Obviously, the lower $T$, the smaller is the range of $t_w$ shown,
because $C^2(t,t_w)$ does not reach 0.2 for all the $t_w$ in our
simulation time.  The errors in $a(t_w)$ and $b(t_w)$ have been
calculated with a Jack-Knife procedure.  The coefficients seems to
have a nice linear behavior in $t_w^{-0.08}$. One thus conclude that,
unless a drastic change arise in the behavior of $a(t_w)$ and
$b(t_w)$, the slope $b(t_w)$ will not vanish asymptotically. This
implies that $C_\mathrm{link}(t,t_w)$ should age as $C^2(t,t_w)$ does.
One must acknowledge, however, that the conclusion is less sound for
$T=0.8$ than for the lower temperatures. In fact, for $T=0.8$ the
infinite-volume extrapolation of $a(t_w)$ (see bottom part
of Fig.~\ref{FIGCOEF3D}) is quite close to the large time-limit of
$C_\mathrm{link}(t_w,t_w)$, as TNT would predict. For the lower temperatures this
extrapolations are clearly different.  The data in four-dimensions
support as well the RSB prediction (see the top part of
Fig.\ref{FIGCOEF4D}).

\section{Conclusions}\label{SECTCONCLUSIONS}

In this work, we report the results of a large scale Monte Carlo
simulation of the heath-bath dynamics of the three dimensional
Edwards-Anderson model with binary coupling at temperatures $T=0.7
T_\mathrm{c}$, $0.53 T_\mathrm{c}$ and $0.44 T_\mathrm{c}$ (see
table~\ref{SIM_TABLE}). The long times achieved and the large lattices
studied ($60^3$), have been made possible by the SUE machine. In
addition, shorter but more precise\footnote{Due to the larger number
of simulated samples.}  simulations are presented for the Metropolis
dynamic of the same model in 3D, 4D and the infinite-dimensional
Viana-Bray model.

For the spin correlation-function in $D=3$ and $D=4$, we find that an
slightly modified power law (see Eq.(\ref{LAECUACION})), well
describes the data for $t>t_w$. This formula is identical to the
cross-over like parametrization (\ref{EQRIEGER}) proposed by Rieger et
al.~\cite{RIEGERFIRST,RIEGER}. However, in the regime $1\ll t\ll t_w$
a different behavior is observed. A numerical procedure is proposed
for the study of the different time-sectors believed to exist in
spin-glasses dynamics~\cite{REV-DYN}. Indeed, a time sector with
characteristic exponent $\mu=0.6$ is observed for the first time, we
believe, although to firmly establish this result will require very
precise simulations at still larger $t_w$.  In spite of this, the very
slow $t_w$ evolution of $A(t_w)$ and $\alpha(t_w)$ (see
Eq.(\ref{LAECUACION})), makes us to believe that in thermoremanent
magnetization measurements (usually restricted to the range $10^{-2}
t_w < t < 10^2 t_w$), a perfect full-aging behavior occurs, in
agreement with a recent experiment~\cite{RODRIGUEZ}, and mild
disagreement with older ones~\cite{SITGES96} (see
also~\cite{MEMORY}). Although it is somehow disappointing that the
study of the (in)existence of more than two-time sectors in spin-glass
dynamics should be restricted to simulations, we still think that it
should be feasible. Nevertheless, one can always hope that new
experimental techniques and protocols will be eventually able to
explore this time-regime.

In infinite-dimensions, or at least for the $z=6$ Viana-Bray model, we
have found a different scaling. The decay of the spin-correlation
function is not a power law. Moreover, the limiting functional-form
Eq.(\ref{LIMITING}) is {\em not} dynamically ultrametric. Although
clearly more work is needed to establish this result, it suggests that
dynamical ultrametricity is maybe not such an interesting property as
previously thought.

We have considered as well the aging properties of the link-overlap
and the link correlation-function, both in three dimensions and in
4D. We have concluded that, unless a drastic change in the dynamical
properties arises for $t_w$ larger than the here studied, the link
correlation-function should age precisely in the same way as the spin
correlation-function. This is in plain disagreement with the droplet
and the TNT pictures of the spin-glass phase. However, one must
acknowledge that the data at the highest studied temperature in three
dimensions ($T=0.7 T_\mathrm{c}$) are not incompatible with the TNT
picture.

\ack
We are indebted with L.A. Fern\'andez and J.J. Ruiz-Lorenzo for
discussions. We thank the Spanish MCyT for financial support through 
research contracts FPA2001-1813,FPA2000-0956,BFM2001-0718 and PB98-0842. 
V.M.M. is a Ram\'on y Cajal research fellow (MCyT) and S.J. is a DGA fellow.\\


\begin{thebibliography}{99}
\bibitem{EXPBOOK} J. A. Mydosh, {\em Spin Glasses: an
Experimental Introduction} (Taylor and Francis, London 1993).

\bibitem{BOOKS} K. Binder and A. P. Young, Rev. Mod. Phys. {\bf 58},
801 (1986); K. H. Fisher
and J. A. Hertz, {\em Spin Glasses} (Cambridge University Press,
Cambridge U.K. 1991)

\bibitem{BOOKRSB} M. M\'ezard, G. Parisi and M. A. Virasoro, {\em Spin
Glass Theory and Beyond} (World Scientific, Singapore 1987);
E. Marinari, G. Parisi, F. Ricci-Tersenghi, J. J. Ruiz-Lorenzo and
F. Zuliani, J. Stat. Phys. {\bf 98}, 973 (2000).

\bibitem{YOUNGBOOK}
{\em Spin Glasses and Random Fields}, edited by A. P. Young. World
Scientific (Singapore, 1997).

\bibitem{AGINGDISCOVER} R.V. Chamberlin, M. Hardiman and R. Orbach,
J. Appl. Phys. {\bf 52,} 1771 (1983); L. Lundgren, P. Svelindh,
P. Norblad and O. Beckman, Phys. Rev. Lett. {\bf 51,} 911 (1983) and
J. Appl. Phys. {\bf 57,} 3371 (1985).

\bibitem{SITGES96} E. Vincent, J. Hamman, M. Ocio, J. P. Bouchaud and
L.F. Cugliandolo in {\em Complex behaviour of glassy systems,\/}
ed. M. Rubi, Springer-Verlag Lecture Notes in Physics {\bf 492,} 184
(1997) (cond-mat/96072224).

\bibitem{MEM-AND-REJ} Ph. Refregier, E. Vincent, J. Hamman and
M. Ocio, J. Physique (France) {\bf 48,} 1533 (1987); K. Jonason,
P. Norblad, E. Vincent, J. Hamman and J. P. Bouchaud, Eur. Phys. J. B
{\bf 13,} 99 (2000); J.P. Bouchaud, V. Dupuis, J. Hamman and
E. Vincent, Phys. Rev. B {\bf 65,} 024439 (2002).

\bibitem{REV-DYN} For a review see J.P. Bouchaud, L.F. Cugliandolo, J. Kurchan
and M. M\'ezard, {\em Out of equilibrium dynamics in Spin-Glasses and
other Glassy Systems} in~\cite{YOUNGBOOK}.


\bibitem{RIEGER} J. Kisker, L. Santen, M. Schreckenberg and H. Rieger,
Phys. Rev. B {\bf 53,} 6418 (1996).  For a review see H. Rieger, in
{\em Annual Reviews of Computational Physics II} (World Scientific
1995, Singapore) p. 295.


\bibitem{JJ} J.J. Ruiz-Lorenzo, {\em Low temperature properties of
Ising spin glasses: (some) numerical simulations.} To appear in
"Advances in Condensed Matter and Statistical Mechanics",
Ed. E. Korutcheva and R. Cuerno. To be published by Nova Science
Publishers, preprint cond-mat/0306675.


\bibitem{FRANZ} S. Franz, M. M\'ezard, G. Parisi, L. Peliti,
Phys. Rev. Lett. {\bf 81}, 1758 (1998); J. Stat. Phys. {\bf 97}, 459 (1999).

\bibitem{FDT} E. Marinari, G. Parisi, F. Ricci-Tersenghi and
J. J. Ruiz-Lorenzo, J. Phys. A {\bf 31}, 2611 (1998); S. Franz and
H. Rieger, J. Stat. Phys. {\bf 79}, 749 (1995).


\bibitem{FDTEXP} D. H\'erisson and M. Ocio, Phys. Rev. Lett. {\bf 88},
257202 (2002).

\bibitem{DROPLETS}
W. L. McMillan,
  J. Phys. C {\bf 17}, 3179 (1984).
  A. J. Bray and M. A. Moore, 
  in {\it Heidelberg Colloquium on Glassy Dynamics}, 
  edited by J. L. Van Hemmen and I. Morgenstern 
  (Springer Verlag, Heidelberg, 1986), p. 121.
  D. S. Fisher and D. A. Huse,
  Phys. Rev. Lett. {\bf 56}, 1601 (1986);
  Phys. Rev. B {\bf 38}, 386 (1988).

\bibitem{TNT} M. Palassini and A. P. Young, Phys. Rev. Lett. {\bf 85},
3017 (2000).

\bibitem{APE} M. Picco, F. Ricci-Tersenghi and F. Ritort, Phys. Rev. B {\bf 63,} 174412 (2001). 


\bibitem{MEMORY}  L. Berthier and J. P. Bouchaud, Phys. Rev. B {\bf
  66}, 054404 (2002).

\bibitem{BRAY} A.J. Bray, Adv. Phys. {\bf 43,} 357 (1994).

\bibitem{RODRIGUEZ} G.F. Rodriguez, G.G. Kenning and R. Orbach,
Phys. Rev. Lett. {\bf 91,} 037203 (2003).

\bibitem{RIEGERFIRST}
H. Rieger, J. Phys. A{\bf 26}, L615 (1993). 

\bibitem{CLU} J. Pech, A. Taranc\'on and C. L. Ullod,
Comput. Phys. Commun. {\bf 106}, 10 (1997).

\bibitem{RUCLU} J. J. Ruiz-Lorenzo and C. L. Ullod.
Comput. Phys. Commun. {\bf 125}, 210 (2000).
 
\bibitem{SUE}
A. Cruz, J. Pech, A. Taranc\'on, P. T\'ellez, C. L. Ullod, C. Ungil,
Comput. Phys. Commun. {\bf 133,} 165 (2001).

\bibitem{LENGTH} E. Marinari, G. Parisi, F. Ricci-Tersenghi and 
J.J. Ruiz-Lorenzo, J. Phys. A {\bf 33}, 2373 (2000).

\bibitem{SUEFSS} H.G. Ballesteros, A. Cruz, L.A. Fernandez,
V. Martin-Mayor, J. Pech, J.J. Ruiz-Lorenzo, A. Tarancon, P. Tellez,
C.L. Ullod and C. Ungil, Phys. Rev. B {\bf 62,} 14237 (2000).

\bibitem{GIORGIOMARC} M. M\'ezard and G. Parisi, Eur. Phys. J. B {\bf
20,} 217 (2001).

\bibitem{JJ2002} E. Marinari, G. Parisi and J.J. Ruiz-lorenzo, J. of Phys. A: Math. and Gen. {\bf 35,} 6805 (2002).

\bibitem{JJa} 
E. Marinari, G. Parisi and J.J. Ruiz-lorenzo, Phys. Rev. B
{\bf 58,} 14852 (1998); E. Marinari, G. Parisi F. Ritort and 
J.J. Ruiz-lorenzo, Phys. Rev. Lett. {\bf 76,} 843 (1996).

\end{thebibliography}
\end{document}